\documentclass[pra,amssymb,amsmath,longbibliography,twocolumn]{revtex4-1}

\usepackage{color,braket,tikz}

\newcommand{\bla}{\color{black}}

\newcommand{\be}{\begin{equation}}
\newcommand{\ee}{\end{equation}}

\newtheorem{defn}{Definition}

\begin{document}

\title{Maximally nonlocal subspaces} \author{Akshata Shenoy H.}
\email{akshata.shenoy@etu.unige.ch} \affiliation{Group of Applied
  Physics, University of Geneva, CH-1211 Geneva, Switzerland}
\author{R.  Srikanth} \email{srik@poornaprajna.org}
\affiliation{Poornaprajna Institute of Scientific Research, Bengaluru,
  India}

\begin{abstract}
A nonlocal subspace $\mathcal{H}_{NS}$ is a subspace within the
Hilbert space $\mathcal{H}_n$ of a multi-particle system such that
every state $\psi \in \mathcal{H}_{NS}$ violates a given Bell
inequality $\mathcal{B}$.  Subspace $\mathcal{H}_{NS}$ is maximally
nonlocal if each such state $\psi$ violates $\mathcal{B}$ to its
algebraic maximum.  We propose ways by which states with a stabilizer
structure of graph states can be used to construct maximally nonlocal
subspaces, essentially as a degenerate eigenspace of Bell operators
derived from the stabilizer generators.  Two cryptographic applications-- to quantum information splitting and quantum subspace certification-- are discussed.
\end{abstract}

\maketitle

\section{Introduction}

Quantum nonlocality is a fundamental quantum feature that demonstrates
that  quantum  mechanics   can't  be  explained  by   a  local  theory
\cite{clauser1969proposed}. It forms the basis for nonclassical tasks,
such  as   secure  key   distribution  with   uncharacterized  devices
\cite{barrett2005no,    *acin2007device,    *vazirani2014fully}    and
device-independent random number generation \cite{colbeck2011private}.
For a  recent review, see \cite[Section  IV]{brunner2014bell}.  It has
been  extensively studied  over the  last 50  years, with  its various
aspects studied  in the bipartite and  multipartite scenario, involving
dichotomic  or  many-valued  measurements.

In particular,  it is  known that the  set $\mathcal{Q}$  of bipartite
quantum correlations is (strictly?)  contained in the set of bipartite
correlations obtained based on the assumption that the local operators
of two observers commute \cite{fritz2012, navascues2012tsirelson}, and
strictly   contained  in   the   set   of  no-signaling   correlations
\cite{PR94}.  As  the set  $\mathcal{L}$ of  local correlations  for a
given finite number  of inputs and outputs is convex,  it follows from
the hyperplane  separation theorem that given  element ${\bf p}^\prime
\notin \mathcal{L}$, there is a  correlation inequality, linear in the
inputs and  outputs, that is  satisfied by  all elements ${\bf  p} \in
\mathcal{L}$ but violated by ${\bf p}^\prime$-- i.e., a witness of the
nonlocality of  ${\bf p}^\prime$.  These inequalities  are called {\it
  Bell inequalities}.  (The corresponding inequalities for the quantum
set ${\mathcal  Q}$ are called \textit{Tsirelson  inequalities}). Bell
inequalities  that  are   \textit{facet  inequalities}--  i.e.,  tight
witnesses--  characterize  $\mathcal{L}$  minimally.   The  facets  of
$\mathcal{L}$ can be determined by  computer codes, but in general the
problem  of determining  whether a  correlation is  local in  the Bell
scenario  with   dichotomic,  multiple  inputs,  is   hard  (in  fact,
NP-complete).

Bell inequalities  for the  bipartite case have  been extended  to the
$n$-particle  situation   \cite{mermin1990extreme,  *ardehali1992bell,
  zukowski2002bell},  which   can  form   the  basis   for  witnessing
multipartite  entanglement   without  assumptions   about  measurement
devices   or  underlying   dimension  \cite{bancal2011device}.    Bell
inequalities  to  witness  $k$-partite   nonlocality,  and  thus  also
$k$-partite entanglement, are known \cite{liang2015family}.

Here, we  shall be  concerned with another,  quite distinct  aspect of
quantum  multipartite nonlocality:  namely, finding  Bell inequalities
that are, for certain given measurements settings, violated equally by
any pure state  in a subspace, called  the \textit{nonlocal subspace}.
Consequently, any superpositions or mixtures of these pure states also
violate the inequality to the same extent.  In a sense, the concept of
a nonlocal  subspace generalizes  the idea  of a  nonlocal state  to a
subspace.

To  the  best of  our  knowledge,  the  problem of  characterizing  or
identifying nonlocal  subspaces hasn't been studied  before.  Here, we
will show  how the  stabilizing properties  of graph  states naturally
conduce  to  the  construction  of quantum  nonlocal  subspaces.   The
subspaces we  construct are \textit{maximally} nonlocal,  in the sense
that the  Bell-type inequality is  violated to its  algebraic maximum.
In  addition to  their  theoretical interest,  nonlocal subspaces  are
experimentally interesting  because they  can be  demonstrated readily
using  practically the  same  setup  used for  tests  of violation  of
Bell-type inequalities.

Furthermore,  their structure  makes them  amenable to  application in
certain quantum  cryptographic tasks, among them,  quantum information
splitting (QIS)  and quantum  subspace certification.  QIS  requires a
quantum state  to be  teleported over  an entangled  state distributed
among various parties. The nonlocal subspace of $n$ particles provides
a natural subspace  in which to encode the quantum  secret (not unlike
the encoding of an unknown state  in a quantum error correcting code),
such that  the nonlocality serves as  the basis to test  security.  We
discuss  later  below  illustrative  examples  that  underscore  these
cryptographic applications.

The  plan of  article is  as  follows.  In  Section \ref{sec:def},  we
formally define nonlocal subspaces and  their maximal kind. In Section
\ref{sec:graph}, we  briefly review  graph states and  how Mermin-Bell
type   inequalities  can   be  constructed   for  them.    In  Section
\ref{sec:belldeg}, we  point out  how to construct  maximally nonlocal
subspaces using graph states, which essentially reduces to the problem
of finding  such an  inequality corresponding to  a \textit{degenerate
  Bell  operator}.   Examples  where  the  degeneracy  can  be  easily
identified are pointed out in  Sections \ref{sec:LC} (case of a linear
cluster  state) and  \ref{sec:cogen} (case  of common  generators).  A
realization  of the  method  for stabilizer  quantum error  correcting
(QEC) codes is given in  Section \ref{sec:qec}, with specific examples
presented in  Section \ref{sec:5} (the  5-qubit QEC code)  and Section
\ref{sec:7}  (Steane code). Finally, we
present  our  conclusions  and   discussions,  with  potential  future
directions, in Section \ref{sec:conclu}.

\section{Nonlocal subspaces \label{sec:def}}

A \textit{measurement  setting} in a  multipartite Bell scenario  is a
set  of given  input choices  of various  observers (say,  Alice, Bob,
Charlie  {\it et  al.})   in an  experiment to  test  whether a  state
violates a Bell inequality.  For example, in an experiment to test the
violation  of the  CHSH  inequality, a  measurement  setting could  be
$\{\{X,  Z\}, \{\frac{1}{\sqrt{2}}(X  \pm  Z)\}\}$,  i.e., that  Alice
chooses one of the Pauli observables  $X$ and $Z$, and Bob chooses one
of $\frac{1}{\sqrt{2}}(X \pm Z)$.

\begin{defn}[Nonlocal subspace] Suppose $\mathcal{H}_n$ is the Hilbert
space of $n$-qubits, and subspace $\mathcal{G} \subset \mathcal{H}_n$,
such that  any every state  $\psi \in  {\mathcal G}$ violates  a given
Bell inequality $\langle{\mathcal B}\rangle \le  L$ to the same degree
for the given measurement setting.  Then, ${\mathcal G}$ is a nonlocal
subspace.
\end{defn}

Here, ${\mathcal  B}$ is the  Bell operator and $L$  the local-realism
bound.  It is  important to stress that this  definition requires that
the  measurement setting,  degree and  Bell inequality  should be  the
same. Otherwise,  the set of  all pure  entangled states would  form a
nonlocal set  in the  sense that  any such state  will violate  a Bell
inequality to some degree for a suitable choice of measurement setting
\cite{gisin1991bell}.

The basic idea here is that  if two distinct states $\ket{\psi_a}$ and
$\ket{\psi_b}$  span  a  nonlocal subspace,  such  that  $\bra{\psi_a}
\mathcal{B} \ket{\psi_b}  = \bra{\psi_b}\mathcal{B}\ket{\psi_b} \equiv
C  > L$,  then  without  any further  calculation,  we  know that  any
superposition  $\alpha\ket{\psi_a}  +  \beta \ket{\psi_b}$  will  also
violate inequality $\langle\mathcal{B}\rangle\le L$ to the same degree
$C$ for a certain fixed measurement setting.

The singlet  state $\ket{\psi^-} \equiv  \frac{1}{\sqrt{2}}(\ket{01} -
\ket{10})$  violates  the  CHSH inequality  $\langle  \mathcal{B}_{\rm
  CHSH} \rangle \equiv |\langle A_1B_1\rangle + \langle A_1 B_2\rangle
- \langle A_2 B_1\rangle + \langle  A_2 B_2\rangle| \le 2$ by reaching
the Tsirelson  bound of $2\sqrt{2}$.  That this is  not a member  of a
nonlocal subspace can be shown by  seeing that any neighoring state to
$\ket{\Psi^-}$  violates  the  inequality  to a  lesser  degree,  that
directly depends on the fidelity.

Specifically,        let        $\ket{\psi(\theta,\phi)}        \equiv
\cos(\theta/2)\ket{01} -  e^{\iota\phi}\sin(\theta/2)\ket{10})$. Then,
one   finds   that   $\langle   \mathcal{B}_{\rm   CHSH}   \rangle   =
2\sqrt{2}F(\theta,\phi)$,  where  $F(\theta,\phi)$   is  the  fidelity
between the states $\ket{\Psi^-}$ and $\ket{\psi(\theta,\phi)}$. Here,
the  fidelity  between states  $\rho_1$  and  $\rho_2$ is  defined  by
$\frac{1}{2}{\rm Tr}\left(\sqrt{(\rho_1 - \rho_2)^2}\right)$.

In general,  the no-signaling bound  on a Bell inequality  exceeds the
quantum (or, Tsirelson)  bound. For example, for  the CHSH inequality,
the Tsirelson bound is $2\sqrt{2}$,  whereas the no-signaling bound is
4  \cite{PR94}.    However,  for   ``all-or-nothing''  type   of  Bell
inequalities (discussed  below), based on a  logical contradiction \`a
la GHZ  \cite{GHZ89}, the quantum  and no-signaling bounds can  be the
same, being equal to the algebraic maximum (the number of terms in the
Bell expression). In such a case, one can consider a strengthened form
of a nonlocal subspace, as defined below.

\begin{defn}[Maximally nonlocal subspace (MNS).] 
Suppose $\mathcal{G}  (\subset \mathcal{H}_n)$ is a  nonlocal subspace
associated with  the Bell  inequality $\langle{\mathcal  B}\rangle \le
L$, such  that any every  state $\psi  \in {\mathcal G}$  violates the
Bell  inequality to  its algebraic  maximum for  the same  measurement
setting.  Then, ${\mathcal G}$ is a maximally nonlocal subspace.
\end{defn}

Following  an  introduction to  graph  states,  we shall  discuss  how
maximally nonlocal  subspaces can  be naturally constructed  for graph
states, essentially by exploiting their stabilizer structure.

\section{Graph states and Bell inequalities: a roundup \label{sec:graph}}

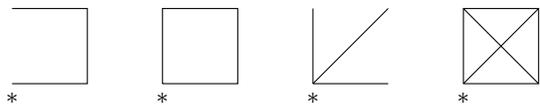
\begin{figure}[h]
\begin{tikzpicture}[scale=1]
\draw (0,0) node[below] {$\ast$} -- (1,0) -- (1,1) -- (0,1);
\draw (2,0) node[below] {$\ast$}-- (3,0) -- (3,1) -- (2,1) -- (2,0);
\draw (4,0) node[below] {$\ast$}-- (5,0);
\draw (4,0) -- (5,1);
\draw (4,0) -- (4,1);
\draw (6,0) node[below] {$\ast$}-- (7,0) -- (7,1) -- (6,1) -- (6,0);
\draw (6,0) -- (7,1);
\draw (6,1) -- (7,0);
\end{tikzpicture}
\caption{Graphs LC$_4$ (linear cluster), RC$_4$ (ring cluster), ST$_4$
  (star topology, rooted at  vertex $\ast$), FC$_4$ (fully connected).
  The last  two are  related by the  graph theoretic  operation called
  ``local complementation'' about vertex $\ast$.}
\label{fig:graph}
\end{figure}

Specifically, graph states are a class of highly entangled multi-qubit
states,  representable  by a  graph  \cite{RB01}.   Given graph  $G  =
(n,\mathcal{E})$, with  $n$ and $E$  being the number of  vertices and
the set of edges, respectively, the graph state $|G\rangle$ is defined
as:
\begin{equation}
|G\rangle = \Pi_{(i,j) \in \mathcal{E}} C(i,j)\ket{+}^{\otimes n},
\label{eq:ising}
\end{equation}
where vertices represent spin systems and edges $C(i,j)$ represent the
controlled-phase  gate  between  qubits  $i$ and  $j$,  which  can  be
realized  using  Ising  interactions. Figure  \ref{fig:graph}  depicts
various graph types with $n=4$.

For  $1\le  j  \le  V$, define  mutually  commuting  local
observables (stabilizers):
\begin{equation}
g_j = X_j\bigotimes_{k \in \textbf{N}(j)}Z_k.
\label{eq:generator}
\end{equation}
where $\textbf{N}(j)$  denotes the  neighborhood of vertex  $j$, i.e.,
the set of  vertices having an edge with vertex  $j$.  The graph state
$\ket{G}$ is simultaneously  the +1 eigenstate of  the $n$ stabilizers
$g_i$:
\begin{equation}
\forall_i ~g_i\ket{G}=\ket{G}.
\label{eq:graphstate}
\end{equation}
Any  graph  state  is  equivalent  a stabilizer  state,  up  to  local
rotations  \cite{nest2004graphical}.  The  set of  all $2^n$  possible
products  (denoted  $h_k$) of  the  generators  $g_i$ forms  a  group,
$\mathcal{S}$, called  the stabilizer.  Obviously, the  graph state is
stabilized by all elements $h_k \in \mathcal{S}$.

A complete basis  for the Hilbert space $\mathcal{H}_n$  of $n$ qubits
can be  derived from $\ket{G}$  by all possible local  applications of
Pauli $Z$ to  the $n$ vertices.  This is the  graph state basis, which
consists of  $2^n$ simultaneous  eigenstates of  stabilizer generators
$g_j$:
\begin{equation}
|G_{\bf
x}\rangle   \equiv  |G_{x_1x_2\cdots   x_n}\rangle  =   \bigotimes_j
\left(Z_j\right)^{x_j}|G_{000\cdots0}\rangle,
\end{equation}   
where $x_j \in \{0,1\}$ and $\ket{G_{000\cdots0}} \equiv \ket{G}$.  It
can be shown that
\begin{equation}
g_j|G_{x_1x_2\cdots
x_n}\rangle = (-1)^{x_j}|G_{x_1x_2\cdots x_n}\rangle.
\label{eq:eigen}
\end{equation}
We define the  \textit{syndrome} of a graph basis state  by the string
$((-1)^{x_1},  (-1)^{x_2}, \cdots,  (-1)^{x_n}) \in  \{\pm1\}^{\otimes
  n}$, which uniquely fixes the graph basis state in the graph basis.

Among various  applications of graph  states we  may count the  use of
cluster   states  in   measurement-based   quantum  computing   (MBQC)
\cite{RB01,  RBB+03,   mantri2017universality}  and   verifiable  MBQC
\cite{hayashi2015verifiable,  markham2018simple}.   Brickwork  states,
which are graph states with the underlying graph being a ``brickwork''
and which require only $X,Y$-plane measurements (rather than arbitrary
$SU(2)$ measuremens) constitute a basic resource for delegated quantum
computation,   specifically   universal  blind   quantum   computation
\cite{broadbent2009universal}.  Graph  states can be used  for quantum
secret  sharing or  quantum information  splitting \cite{MP08,  MP208,
  MS08,  KFM+10,  MMM12,  MMM12,   SSSgr},  quantum  error  correction
\cite{SW01}  and  quantum metrology  \cite{markham2018simple}.   

Graph      states      have     been      realized      experimentally
\cite{KSW+05,LZG+91,bell2014experimental,      bell2014experimental+}.
Their robustness in the  presence of decoherence \cite{HDB05} enhances
their practical value.

As highly entangled states, not surprisingly, graph states show a rich
variety of nonlocal correlations  through the violation of Mermin-type
inequalities \cite{mermin1990extreme} based on stabilizer measurements
\cite{SAS+05,    GTH+05,    cabello2008mermin}   generating    perfect
correlations of GHZ type \cite{GHZ89},  and also through violations of
Bell-Ardehali    inequalities    \cite{ardehali1992bell}   based    on
non-stabilizer measurements \cite{GC08, TGB06}.

  
From any subset of the stabilizer $\mathcal{S}$, we can construct
the operator:\bla
\begin{equation}
\mathcal{B} \equiv \sum_{k=1}^m h_k,
\label{eq:hj}
\end{equation} 
where $m  \le 2^n$, and  the $h_k$'s  are any $m$  distinct stabilizer
elements. Since  the $g_j$'s  are tensor products  of local  (in fact,
Pauli) operators, therefore $h_k$'s are  also tensor products of Pauli
operators.  In view of Eq. (\ref{eq:graphstate}):
\begin{equation}
\mathcal{B}|G\rangle =  m|G\rangle.
\label{eq:ev}
\end{equation}
Let $q$  denote the  largest number of  $h_k$'s in  Eq.  (\ref{eq:hj})
that can assume a positive  value ($+1$) under a local-realistic value
assignment to the individual Pauli  operators. 

In the classical world, each property of each particle can be assigned
values independently of the settings  of other particles, and we would
expect  $q=m$. However,  as we  shall illustrate  by a  simple example
below, in quantum  theory on account of the  non-commutativity and the
intransitivity of commutativity of observables, for certain choices of
operators   $h_k$   in  Eq.   (\ref{eq:hj})   one   may  encounter   a
Greenberger-Horne-Zeilinger    (GHZ)    type   \cite{GHZ89}    logical
contradiction, making  $q$ strictly  smaller than $m$.   Therefore, if
for a  given choice of $m$  operators $h_k$, $q<m$, then  the operator
defined by Eq. (\ref{eq:hj}) constitutes a Bell operator, for which we
can write down a Bell inequality (BI) of the type:
\begin{equation}  
\langle \mathcal{B} \rangle \le \mathcal{L} \equiv 2q-m,
\label{eq:bell}
\end{equation}
which is  violated to its algebraic  maximum (of $m$) by  the relevant
graph state.  We note that  the set  of $m$ stabilizer  elements which
form the  GHZ-type contradiction is  not unique. There can  be several
such  contradictions  derived  from other  subsets  of  $\mathcal{S}$,
leading   to  different   Bell  operators.    Indeed,  the   full  set
$\mathcal{S}$  will lead  to a  Bell  operator, as  noted below.   The
degree    of    violation    of    BI    may    be    quantified    by
$\mathcal{D}=\frac{m}{2q-m}$, which  would be  the relevant  figure of
merit  that  determines  resistence  of the  violation  to  noise  and
detection loophole.  In Eq. (\ref{eq:bell}), there may not be an obvious pattern that allows us to compute $q$. However, it can be  determined by straightforward computer search, by assigning values $\pm1$ to the (at most) three variables $ X, Y$ and $ Z $ for each of the $ n $ qubits, i.e., by searching through (at most) $ 3^{2n} $  possibilities.

Any graph state  violates a BI, which can be  shown using an inductive
argument  \cite{GTH+05}.   Extending this  argument,  the  sum of  all
stabilizer elements  $h_k$ is  a Bell operator,  though not  a maximal
one. In fact:
\begin{equation}
2^{-n}\sum_{k=1}^{2^n} h_k = \ket{G}\bra{G},
\end{equation}
which  is  easily  verified.    There  are  $2^{2^n}$  potential  Bell
operators of  the type (\ref{eq:hj}).  For  $3 \le n \le  6$, they are
fully  characterized   into  14   equivalent  classes  (up   to  local
rotations). Among them are the  multiqubit GHZ states, which correspond
to the star graph \cite{cabello2008mermin}.

Let us  consider a simple example  of a Mermin inequality  for a graph
state, in the case of the linear cluster state LC$_4$:
\begin{equation}
|G\rangle =  \frac{1}{2}\left(|{+}0{+}0\rangle + |{+}0{-}1\rangle
+ |{-}1{-}0\rangle + |{-}1{+}1\rangle\right),
\label{eq:lc4}
\end{equation}
stabilized by generators: $g_1 \equiv X_1Z_2, g_2 \equiv Z_1X_2Z_3,
g_3 \equiv Z_2X_3Z_4$ and $g_4 \equiv Z_3X_4$.

One  constructs a  contradiction  in  a manner  analogous  to the  GHZ
argument \cite{GHZ89}, which is  based on perfect (``all-or-nothing'')
correlations. Consider the 4 stabilizing operators:
\begin{equation}
\begin{array}{lccccc}
~~g_1g_3 & = +X & I & X & Z & \rightarrow +1, \\
~~g_2g_3 & = +Z & Y & Y & Z & \rightarrow +1, \\
~~g_1g_3g_4 & = +X & I & Y & Y & \rightarrow +1, \\
-g_2g_3g_4 & = +Z & Y & X & Y & \rightarrow -1,
\end{array}
\label{eq:logcon}
\end{equation}
Each column has  two copies of a Pauli operator,  meaning that under a
local-realistic assignment  of value  $+1$ or  $-1$ to  the individual
Pauli operators, the column product is 1.  But, the product on the RHS
is $-1$, leading to a contradiction.  Therefore, the sum
\begin{align}
\mathcal{B} =  XIXZ + ZYYZ +  XIYY -
ZYXY
\label{eq:sca05bell}
\end{align}
provides a Bell operator of the Mermin type.

By design,  $\langle G|\mathcal{B}|G\rangle = 4$  for $\mathcal{B}$ in
Eq.   (\ref{eq:sca05bell}), the  number of  summands $m$  in the  Bell
operator.  On the  other hand, the above  contradiction argument shows
that   only   3   terms    in   Eq.    (\ref{eq:sca05bell})   can   be
local-realistically   made  positive,   so  that   $q=3$.   From   Eq.
(\ref{eq:bell}), the local bound $\mathcal{L} = 2q-m= 2$. We thus have
the Bell-type inequality
\begin{equation}
\langle \mathcal{B} \rangle \le 2,
\label{eq:sca05bell+}
\end{equation}
for the Bell operator in (\ref{eq:sca05bell}).

For a  large graph state,  $q$ can be  derived by computer  search.  A
helpful tip here  is that the local-realistic  value assignment scheme
may be assumed to assign $Z=+1$ \cite{GTH+05}. Another tip is that the
value of $q$  is invariant under local complementation.   Thus, in the
case  of  completely  connected  graphs   and  star  graphs,  we  have
$\mathcal{B}$(ST$_n$)    =   $\mathcal{B}$(FC$_n$)    for   a    given
$\mathcal{B}$ (see Figure \ref{fig:graph}).

\section{Bell-degeneracy \label{sec:belldeg}}

Given  any graph  basis $\ket{G^\prime}$, let  $\hat{g}_j$ denote
the  eigenvalue   of  generator  $g_j$,  i.e.,   $g_j\ket{G^\prime}  =
\hat{g}_j\ket{G}_j$.  Since all the generators $g_j$ commute with each
other, and $\ket{G^\prime}$  is a joint eigenstate of theirs, it  is 
readily seen by direct substitution
that any sum  of products of $g_j$'s acting on
$\ket{G^\prime}$ equals the corresponding sum of products
of $\hat{g}_j$'s multiplying $\ket{G^\prime}$. That is,
\begin{align}
(g_{\alpha_1}g_{\alpha_2}&\cdots g_{n_\alpha} +
g_{\beta_1}g_{\beta_2}\cdots g_{n_\beta} +
\cdots )\ket{G^\prime} \nonumber \\
&=  (\hat{g}_{\alpha_1}\hat{g}_{\alpha_2}\cdots \hat{g}_{n_\alpha} +
\hat{g}_{\beta_1}\hat{g}_{\beta_2}\cdots \hat{g}_{n_\beta} +
\cdots )\ket{G^\prime}.
\end{align}
In particular, this means that the Bell operator can be replaced
by the corresponding function of the respective eigenvalues:
$$
\mathcal{B}(g_1, g_2, \cdots, g_n)\ket{G^\prime} 
= \mathcal{B}(\hat{g}_1, \hat{g}_2, \cdots, \hat{g}_n)\ket{G^\prime}
$$
In this light, Eq. (\ref{eq:ev}) can, in view of the form Eq. (\ref{eq:hj}),
be considered as a
set of $m$ constraints (``Bell conditions'') on the graph syndrome:
\begin{equation}
\forall_{k=1}^m \quad h_k(\hat{g}_1,  \hat{g}_2,  \cdots,  \hat{g}_n)=1.
\label{eq:bellcond}
\end{equation}  
That is,  the solution to  these conditions would represent  the graph
syndrome(s) of graph basis state(s) that satisfy Eq.  (\ref{eq:ev}).

If these constraints  don't uniquely fix the graph basis  state to the
graph state $\ket{G}$, then there  will be multiple syndrome solutions
to  Eq.   (\ref{eq:ev}),  making   the  Bell   operator  $\mathcal{B}$
degenerate.

Let $\ket{G_j}$ denote  these multiple graph basis  state solutions to
Eq. (\ref{eq:ev}). \bla  By virtue of linearity,  any normalized state
$\sum_j     \alpha_j    \ket{G_j}$     also     violates    the     BI
$\langle\mathcal{B}\rangle\le\mathcal{L}$  by  reaching its  algebraic
maximum.  Thus,  the span  of these  $\ket{G_j}$'s defines  a subspace
associated  with  maximal  violation.   Accordingly,  this  degenerate
+1-eigenspace  of  $\mathcal{B}$   constitutes  a  maximally  nonlocal
subspace  (MNS), denoted  $\mathcal{H}_{\rm  MNS}$.   Various ways  to
produce Bell degeneracy are exemplified below.

\subsection{Bell degeneracy with LC state \label{sec:LC}}

The    straightforward   method    is    to    solve   the    equation
$\mathcal{B}(\hat{g}_1,\hat{g}_2,\cdots,\hat{g}_n)=m$.  Multiciplicity
of  solutions leads  to Bell  degeneracy, which  may be  determined by
computer search  for large-$n$  graph states.   For the  state LC$_4$,
characterized      by      Eq.      (\ref{eq:sca05bell}),      solving
$\hat{g}_1\hat{g}_3          =           \hat{g}_2\hat{g}_3          =
\hat{g}_1\hat{g}_3\hat{g}_4=\hat{g}_2\hat{g}_3\hat{g}_4=1$,   we  find
solutions   given  by   syndromes  $(\hat{g}_1,\hat{g}_2,   \hat{g}_3,
\hat{g}_4)  \rightarrow (\pm1,\pm1,\pm1,1)$.  Thus, the  corresponding
graph basis states span $\mathcal{H}_{\rm MNS}$.

The first  of these  syndromes correspond  to graph  state $|G\rangle$
given by Eq. (\ref{eq:lc4}), while the other state to
\begin{align}
\ket{G^\prime} &\equiv Z_1Z_2Z_3|G\rangle   \nonumber\\
&=  \frac{1}{2}\left(|{-}0{-}0\rangle      +      |{-}0{+}1\rangle     -
|{+}1{+}0\rangle - |{+}1{-}1\rangle\right).
\label{eq:phi1110}
\end{align}
Any   superposition    in   subspace    $\mathcal{H}_{\rm MNS}$,   namely,
$\alpha|G\rangle   +    \beta|G^\prime\rangle$   also    violates   BI
(\ref{eq:sca05bell+}) to its algebraic maximum of 4.

\subsection{Bell degeneracy via Common generators\label{sec:cogen}} 

In  a  Bell  operator  $\mathcal{B}$, suppose  $l$  $(>1)$  stabilizer
generators $g_1,  g_2, \cdots, g_l$  appear in all the  summands $h_k$
$(1 \le j \le m$).  Then, dim$(\mathcal{H}_{\rm MNS})\ge2^{l-1}$, which is
the  number of  value assignments  to $(\hat{g}_1,  \hat{g}_2, \cdots,
\hat{g}_l)$ consistent with $\hat{g}_1\hat{g}_2\cdots\hat{g}_l = 1$.

As  an  example, consider  the  6-qubit  linear cluster  state  LC$_6$
\cite{cabello2008mermin}:
\begin{align}
\mathcal{B} = 
g_2g_5(I + g_1)(I + g_3)(I + g_4)(I + g_6) \le 4,
\label{eq:sca05bell1}
\end{align}
where $g_1 = X_1Z_2, g_6 =  Z_5X_6 $ and $g_j = Z_{j-1}X_jZ_{j+1}$ for
$j = 2, 3,  4, 5$. In this case, $l=2$ and the  two graph basis states
spanning  $\mathcal{H}_{\rm MNS}$ are  $(\hat{g}_1, \hat{g}_2,  \hat{g}_3,
\hat{g}_4, \hat{g}_5, \hat{g}_6) \rightarrow (1,\pm1, 1, 1, \pm1, 1)$,
with
$$ \ket{G^\prime} = Z_2Z_5\ket{\rm LC_6},$$  being the second state in
addition to  $\ket{G}$ that  violates BI (\ref{eq:sca05bell1})  to its
algebraic maximum.

\section{Bell degeneracy for QEC codes\label{sec:qec}}

Quantum error correcting  (QEC) codes have a  natural association with
MNS.  A  $[[n, k]]$ QEC  code encodes $k$  qubits in $n$  qubits, such
that  the  code  space  is  stabilized  by  $n-k$  commuting  syndrome
operators  $g_j$  \cite{gottesman1997stabilizer}.  Any  Bell  operator
$\mathcal{B}$ formed from these $(n-k)$ generators will obviously have
a  $2^k$-fold degeneracy,  since all  states  in the  code space  will
produce maximal violation, by construction.

\subsection{5-qubit QEC code\label{sec:5}}

As an example,  let $|G_0\rangle$ and $|G_1\rangle$ be  the code words
for the 5-qubit code \cite{BDS+96}, which corrects one arbitrary qubit
error.
\begin{eqnarray}
|G_0\rangle  &=&   \frac{1}{4}  
(-  |00000\rangle   -  |11000\rangle  - |01100\rangle  -  
|00110\rangle \nonumber \\
&-& |00011\rangle  -  |10001\rangle  +
|10010\rangle  +  |10100\rangle   +  |01001\rangle  \nonumber  \\  &+&
|01010\rangle  +  |00101\rangle  +  |11110\rangle +  |11101\rangle  +
|11011\rangle \nonumber \\
&   + &  |10111\rangle    +    |01111\rangle)   \nonumber\\
|G_1\rangle &=& XXXXX|G_0\rangle.
\label{eq:steane}
\end{eqnarray}
where $XXXXX \equiv X^{\otimes 5}$.
The stabilizers are $g_1  =
XYYXI$,  $g_2= IXYYX$,  $g_3 =  ZYIYZ$ and  $g_4=
XYZYX$.

It may be checked that 
\begin{equation}
\mathcal{B}=
g_4g_1(1+g_3) + g_2(g_1+g_3) + g_1 \le 3
\label{eq:5qub}
\end{equation}
constitutes a Mermin inequality  with $m=5$.  Our previous observation
entails  that any  encoded  state in  this QEC  code  will violate  BI
(\ref{eq:5qub}) maximally.   It follows from  Eq.  (\ref{eq:5qub})
that $\hat{g}_2\hat{g_3}=  \hat{g}_2\hat{g_1}= 1$, and  therefore that
$\hat{g}_i$  ($i=1,2,3$) have  the same  sign.  Because  of the  first
summand   in   Eq.     (\ref{eq:5qub}),   $\hat{g_3}=1$   and   thus
$\hat{g}_4=1$.  In other words, the  ``Bell conditions'' fully fix the
code  space, and  there is  no further  degeneracy.  But  this is  not
necessary, as we discuss with the Steane code.

\subsection{Steane QEC code\label{sec:7}}

A  BI  that  can  be  constructed for  the  7-qubit  Steane  QEC  code
\cite{S96}, given by:
\begin{eqnarray}
\mathcal{B} = g_2g_1(1 + g_4 + g_5g_4) + g_5g_3(g_1 + g_2) + g_5 \le 4
\label{eq:bellst}
\end{eqnarray}
where the stabilizer generators for the Steane code are 
$g_1  = IIIXXXX, g_2 =  IXXIIXX, g_3 = XIXIXIX, 
g_4 = IIIZZZZ, g_5 =  IZZIIZZ$ and $g_6 = ZIZIZIZ$. 

Note   that   the  generator   $g_6$   doesn't   appear  in   the   BI
(\ref{eq:bellst}), meaning that  the value assignment $\hat{g}_6$
is unrestricted.  Solving the  ``Bell conditions'' for $\hat{g}_j$ ($1
\le       j       \le       5$)       gives       two       solutions:
$(\hat{g}_1,\hat{g}_2,\hat{g}_3,\hat{g}_4,\hat{g}_5)       \rightarrow
(\pm1,\pm1,\pm1,1,1)$.  For BI (\ref{eq:bellst}), we thus find
$$
\textrm{dim}(\mathcal{H}_{\rm MNS})= 4 \times \textrm{dim(code~space)}
= 8.
$$ Thus, not just states in QEC code space, but other states indicated
by  these  graph  syndromes would  violate  BI  (\ref{eq:bellst})
maximally. Some of these may  correspond to correctible erroneous code
words (when  the Hamming  weight of corresponding  error vector  is at
most 1,  e.g., the graph  basis state  corresponding to $Z_7$)  or not
(e.g., that corresponding to $Z_1Z_2Z_3$).

\section{Applications\label{sec:app}}

An MNS can  be adapted to various applications where  graph states are
used, with  the key extension  that not just  a resource state,  but a
whole resource  subspace is available.  Potential  areas of employment
include   metrology,  $t$-designs   \cite{markham2018simple},  quantum
cryptography,  measurement-based quantum  computing (MBQC)  verifiable
MBQC and universal blind quantum computation.

Here  we shall  consider  its  applications to  two  tasks in  quantum
cryptography.  One of them is QIS,  wherein the secret is encoded in a
nonlocal    subspace   and    then   distributed    among   legitimate
agents.  Because of  entanglement  monogamy, the  effect  of noise  or
eavesdroppers will  be to lower  the degree  of violation of  the Bell
inequality.  This provides an experimental basis to test the integrity
and security of the shared secret.

Another application of our approach  is that of certifying a subspace,
in that the associated Bell inequality is violated maximally by states
in the  subspace alone.  We  discuss below illustrative  examples that
underscore these cryptographic applications.

\subsection{Quantum information splitting \label{sec:qss}}

In  standard  QIS  \cite{SS05},   the  secret  dealer  distributes  an
entangled state with suitable properties, such as a graph state, among
the players. This state acts as  a channel, over which the dealer then
teleports an unknown state $\ket{\psi}$ (the secret). By contrast, our
present  approach  contains   a  twist  to  this   plot,  whereby  the
entanglement channel  will already encode  the secret in the  MNS, and
eventually   the   dealer   simply  teleport   a   suitable   fiducial
(secret-independent)  state,  $\ket{0}$   by  convention,  across  the
channel,  such that  after  the classical  communication from  various
parties,  the  secret  $\ket{\psi}$  is recovered  by  the  designated
player.

In QIS, a  secret dealer distributes an encoded quantum  state among a
certain  number  of players,  so  that  their collaboration  allows  a
designated player  (the recoverer)  to recover  the states.  The basic
idea  can be  illustrated  for  a QIS  scheme  with  the 5-qubit  code
mentioned  above.  Let  Alice (the  secret dealer)  have qubit  1, Bob
qubits 2 and 3, Charlie qubit 4, while Rex (the secret recoverer) have
qubit  5.    The  secret   $\ket{\psi}\equiv\mu\ket{0}+\nu\ket{1}$  is
encoded by Alice in the space $\mathcal{H}_{\rm MNS}$ as $\mu\ket{G_0}
+  \nu\ket{G_1}$,  which Alice  distributes  among  all players.   The
reduced density operator  with each player should  hold no information
about the secret, which may ideally  be an arbitrary qubit state.  For
perfect secrecy, graphs that correspond  to QEC codes \cite{CGL99} are
appropriate, while the  properties of MNS may be used  for testing the
code space.

In the  first step of  the protocol, Alice  measures her qubit  in the
computational basis.   The result  is given  in Table  \ref{tab:1}. In
step 2, Bob  measures in the computational basis, too,  the results of
which  are depicted  in Table  \ref{tab:2}  for the  case where  Alice
obtained $\ket{0}$ in the first step.  In step 3, Charlie measures his
qubit in  the computational basis.  These steps leave the  secret with
Rex's qubit, up to a Pauli  operation.  Clearly, Rex can find out this
Pauli  operator, and  thereby  recover the  quantum  secret, based  on
classical communication from Alice, Bob, Charlie, and can't recover it
without input from even one of them.

\begin{table}
	\begin{tabular}{l|l}
		\hline
		Outcome  & State with Bob, Charlie and Rex\\
		of Alice & \\
		\hline
		$\ket{0}$ & $ \mu\Big(-\ket{0000} - \ket{0011}  - \ket{0110} - \ket{1100}$ \\
		& $+ \ket{0101} + \ket{1001} + \ket{1010}  + \ket{1111} \Big)$ \\
		~ & $+ \nu\Big(\ket{1101} -\ket{0111}
		- \ket{1110} + \ket{1011}$ \\
		&+ $\ket{0010} + \ket{0001} + \ket{1000} + 
		\ket{0100}\Big)$ \\
		\hline
		$\ket{1}$ & $ \mu\Big(-|1000\rangle - |0001\rangle + |0010\rangle + |0100\rangle$ \\
		& $+ |1110\rangle + |1101\rangle + |1011\rangle + |0111\rangle\Big)$ \\
		~ & $+ \nu\Big(\ket{0110\rangle} -
		\ket{0011} - \ket{1111} - \ket{1001}$ \\
		& $- 
		\ket{1100} + \ket{0000} + \ket{1010}
		+ |0101\rangle\Big)$\\
		\hline
	\end{tabular}
	\caption{QIS  using the  5-qubit  code  (\ref{eq:steane}): Outcome  of
		Alice's measurement with corresponding  state left with Bob, Charlie
		and Rex.}
	\label{tab:1}
\end{table}

\begin{table}[h]
	\begin{tabular}{l|l}
		\hline Outcome & State left with \\
		of Bob & Charlie and Rex \\ \hline
		$\ket{00}$ & $\ket{0}(Z\ket{\psi}) + \ket{1}(Y\ket{\psi})$ \\ 
		$\ket{11}$  & $\ket{0}(Z\ket{\psi}) - \ket{1}(Y\ket{\psi})$ \\  
		$\ket{01}$ & $\ket{0}(X\ket{\psi}) + \ket{1}(I\ket{\psi})$ \\ 
		$\ket{10}$  & $\ket{0}(X\ket{\psi}) - \ket{1}(I\ket{\psi})$ \\ 
		\hline
	\end{tabular}
	\caption{Bob's measurement  outcome, and the corresponding  state left
		with Charlie-Rex, up to a global phase, when Alice obtains $\ket{0}$
		in  Table \ref{tab:1}.  Rex can  reconstruct the  state safely  with
		inputs from the  rest.}
	\label{tab:2}
\end{table}

Let  us  consider a  simple  security  scenario  of this  QIS  scheme.
Suppose Eve,  as part of  eavesdropping, attacks  the 4th qubit  of an
encoded state of above 5-qubit QECC,  as part of which she employs the
two-qubit controlled-qubit interaction:
\begin{equation}
U(\theta)  =  \ket{0}\bra{0}\otimes\mathbb{I}  + 
\ket{1}\bra{1}\otimes
\left(  \begin{array}{cc}  \cos\eta  & \sin\eta  \\  \sin\eta  &
-\cos\eta \end{array}\right),
\end{equation}
where $0  \le \eta  \le \pi/2$.   By straightforward  calculation, one
finds  that under  this interaction,  the expectation  values for  the
stabilizing elements are
\begin{equation}
\langle h_m\rangle =
\left\{
\begin{array}{cc}
\cos(\eta) & ~~(m=1,5)\\
1 & ~~(m=2,3,4),
\end{array} \right.
\end{equation}
from which it follows that for BI (\ref{eq:5qub})
\begin{equation}
\langle \mathcal{B} \rangle = 2\cos(\eta) + 3,
\end{equation}
which reaches  the local bound  3 when $\eta=\pi/2$.  Thus,  the basic
idea is that any intervention by Eve diminishes the level of violation
away from maximality. Quite generally, this behavior is related to the
monogamy  of   quantum  entanglement  and  of   nonlocal  no-signaling
correlations.

\subsection{Quantum subspace certification \label{sec:cert}}

Given an unknown system  and uncharacterized measurement devices, some
system  features,  such  as  its dimension  or  entanglement,  may  be
inferrable  from  the  observed measurement  statistics.   Thus,  such
features admit  self-testing \cite{MY98, mayers2004self},  wherein one
makes no assumptions about preparations, channels and measurements.

State tomography or entanglement witnesses also test states, but under
the assumption of trusted preparation and measurement procedures.  The
problem of certifying states requires  a higher level of trust,
where measurements are trusted, but sources and channels aren't.  This
can  be extended  to  a  self-test, essentially  by  showing   robustness against errors in  the measuring instruments.
Typically, a self-test requires the violation of a suitable Bell-type inequality against specific local measurements.

Here, we shall briefly discuss how our approach to identify an MNS can be used to construct a produce to certify the subspace.
This generalizes the problem of certifying a given graph state \cite{mckague2011self}.
Because graph  basis states  form a  complete basis,  stabilizer tests
which admit an  MNS such that dim$(\mathcal{H}_{\rm MNS}) >  1$ can be
used to certify  that the state belongs to the  subspace $H_{\rm MNS}$
in question by  verifying that it maximally (or,  to sufficiently high
degree) violates  the associated  Bell inequality  $\mathcal{B}$.  Two
security  criteria here  are \cite{markham2018simple}:  (Completeness)
that  the   test  accepts  an  ideal   preparation;  (Soundness)  that
acceptance  indicates   sufficient  closeness   to  the   ideal  state
preparation.

We note that the state certified in this way, while guaranteed to be an element
of $H_{\rm MNS}$, may be a pure or mixed.  If, further, a guarantee of
purity  is needed,  a further  component to  self-test purity  must be
added.    In  this
framework, we obtain the certification of a given graph state as a special
case  of subspace  certification,  where  one seeks  an  MNS of unit dimension.  That is, suppose $\ket{G}$ uniquely
violates BI  $\mathcal{B}$ maximally, but  no other graph  basis state
does (an example is discussed below).   Stabilizers  $g_j$  associated  with  $\mathcal{B}$  obviously
accept  $\ket{G}$, which  guarantees completeness.   By virtue  of the
assumed uniqueness, any deviation of the prepared state from $\ket{G}$
will increase chances of rejection, leading to soundness.

By way  of an example:  Ref. \cite{cabello2008mermin} lists BI's  for graph
states  of  various families  with  up  to  6 qubits.   Three  4-qubit
inequalities listed for $\ket{LC_4}$ are:
\begin{subequations}
\begin{align}
	\mathcal{B}_1 &=  (I + g_1)g_2(I + g_3) \le 2 \label{eq:lc4a}\\
	\mathcal{B}_2 &=  (I + g_1)g_2(I + g_3g_4) \le 2 \\
	\mathcal{B}_3 &=  (I + g_1)g_2(g_3 + g_4) \le 2,
\end{align}
\end{subequations}
where $g_1 = X_1Z_2, g_4 = Z_3X_4$ and $g_j = Z_{j-1}X_jZ_{j+1}$ ($j =
2, 3$).  The maximal algebraic and quantum bound are 4 in each case.

By    inspection,  for each of these three inequalities, we    find   that the dimension of the corresponding MNS is 2,  since (corresponding) $(\hat{g}_1, \hat{g}_2, \hat{g}_3,  \hat{g}_4)_1  \rightarrow (1,1,1,\pm1)$  and  $(\hat{g}_1,
\hat{g}_2,   \hat{g}_3,   \hat{g}_4)_2  \rightarrow   (1,1,\pm1,\pm1)$ and $(\hat{g}_1, \hat{g}_2, \hat{g}_3,  \hat{g}_4)_1  \rightarrow (1,\pm1,\pm1,\pm1)$ all violate the corresponding BI maximally.   Therefore, maximal (or
close to maximal) violation of one of these  inequalities can be used to certify the corresponding graph subspace. For example, a (near) maximal violation of inequality (\ref{eq:lc4a}) would indicate the state is a superposition of $\ket{LC_4}$ and $Z_4\ket{LC_4}$.

The 5-qubit state $\ket{GHZ_5}$ \cite{cabello2008mermin}, which is stabilized by the five operators $g_1=X_1Z_2Z_3Z_4Z_5$ and $g_j = Z_1X_j$ ($j = 2,3,4,5$))     is    the     unique    state that maximally violates  
\begin{equation}
g_1(I + g_2)(I+g_3)(I+g_4)(I+g_5) \le 4
\label{eq:ghz5}
\end{equation}
to its algebraic maximum of 16.
Therefore, maximal (or
close to maximal) violation of  inequality Eq. (\ref{eq:ghz5}) can be used to certify the state $\ket{GHZ_5}$.

This  method  of certification  can  be  employed  in the  context  of
verifiable MBQC, allowing this idea  to be extended to fault tolerance
by  having client  (Alice) ask  server (Bob)  for a  suitable resource
graph state (cf.  \cite{fuiji2017verifiable}),  such as the 3D cluster
state     used    in     a    fault-tolerant     topological    scheme
\cite{raussendorf2007topological}.

\section{Conclusions and discussions\label{sec:conclu}}

We  proposed  various ways  by  which  graph  states  can be  used  to
construct maximally nonlocal subspaces,  essentially as the degenerate
eigenspaces of Bell operators  derived from the stabilizer generators.
Applications to  quantum cryptography  were discussed,  in particular,
quantum information splitting and quantum subspace certification.

A future direction would be to extend our approach to develop a method
for creating  nonlocal subspaces for  Bell-Ardehali-type inequalities,
which aren't based on stabilizer measurements but may lead to stronger
violations of the  relevant BI.  Another direction would  be to derive
Svetlichny-type   inequalities    for   graph   states    leading   to
\textit{absolutely nonlocal subspaces} for graph states.  

\acknowledgments  Akshata  Shenoy  H. acknowledges  the  support  from
Federal Commission  for Scholarships for Foreign  Students through the
Swiss      Government      Excellence     Postdoctoral      Fellowship
2016-2017.  R. Srikanth  thanks the  Defense Research  and Development
Organization  (DRDO),  India  for  the support  provided  through  the
project number ERIP/ER/991015511/M/01/1692.

\bibliography{axta}

\end{document}